\documentclass{aa}
\usepackage[dvips]{graphics}
\begin{document}
 
\title{The hard X-ray emission of luminous infrared galaxies}
  
\author{G. Risaliti\inst{1}, R. Gilli\inst{1}, 
R. Maiolino\inst{2}, M. Salvati\inst{2}
}
   
\institute{
Dipartimento di Astronomia e Scienza dello Spazio,
Universit\`a di Firenze Largo E. Fermi 5, I--50125 Firenze, Italy 
(risaliti, gilli@arcetri.astro.it)
\and
Osservatorio Astrofisico di Arcetri, Largo E. Fermi 5,
I--50125 Firenze, Italy (maiolino, salvati@arcetri.astro.it)
}
\authorrunning{Risaliti et al.}
\titlerunning{X-ray properties of LIGs}
\offprints{G. Risaliti}
     
\date{}
     
\thesaurus{03(11.01.2; 11.19.3; 13.09.1; 13.25.2)}
      
\maketitle

\begin{abstract}
We present a study of the hard X-ray properties of a sample that
includes all
the Luminous Infrared Galaxies (LIGs, $\rm L_{IR} > 10^{11}$
L$_\odot$) observed in the 2-10 keV energy band (new and archival data).
We find that
a significant fraction of the
sources optically classified as AGNs do not show any indication of
nuclear activity in the X rays, thus suggesting heavy absorption
along our line of sight.
The absence of strong emission in the 20-200 keV band in a subsample
of LIGs
observed with BeppoSAX suggests that in many cases these sources are
completely Compton thick ($\rm N_H > 10^{25}$ cm$^{-2}$).
From a comparison between the infrared and the X-ray emission we deduce
that the mid-IR emission is absorbed by a lower column density than the
X-ray emission or, alternatively, that the dust-to-gas ratio is lower
than Galactic. We describe a simple model that reproduces the
IR--X correlation by means of mixed AGN and starburst contributions and
we compare the predictions of this model with the observational data at
X-ray and optical wavelengths. Finally, we discuss the biases that affect
the currently available samples of LIGs and briefly analyze a small unbiased
sample, finding that at least 50\% of the sources host a (weak) AGN.
\end{abstract}
	 
	  
\section{Introduction.}
The nature and the energy source of the Luminous Infrared Galaxies
(LIGs) has been a matter of debate since their discovery by 
IRAS, more than 10 years ago. This class of sources is
characterized by a very high luminosity emitted mostly in the 
far--infrared ($\rm L_{IR} > 10^{11}
L_{\odot}$). Such a luminosity is higher than in
normal galaxies ($\rm L_{IR}\sim 10^{10}$ erg s$^{-1}$ 
for a large spiral like the Milky Way or
M31) and therefore an additional source of energy is
required. The infrared emission of LIGs is due to the presence of
a large amount of dust that reprocesses optical and UV
radiation to IR wavelengths. Since the primary emission
is often not directly observable, its origin remains controversial.

Two main mechanisms have been invoked in this context:
an intense starburst activity and/or an Active Galactic Nucleus 
(AGN). It is now widely accepted that in general
both sources can be present in the LIGs, but the
relative contribution of each is still unclear. 

Recent studies with ISO of a sample of ULIGs (Ultra Luminous
Infrared Galaxies, $\rm L_{IR} > 10^{12} L_{\odot}$) indicate that
the bulk of the IR emission is due to starburst activity (Genzel et al.
1998, Lutz et al. 1998). The diagnostic used in these works is mainly
based on mid-infrared spectroscopy: in particular, the policyclic
aromatic hydrocarbon features at 7.7 $\mu$m are claimed to be better
than optical lines at discriminating between starburst and AGN, 
because of the lower extinction.

On the other side, numerous X-ray observations performed in the last few
years, which will be in part reviewed in this paper, suggest that an AGN
contribution to the observed IR luminosity is common.

Resolving this issue would be important not only
for the understanding of the LIG phenomenon,
but also to assess the contribution of the LIGs
to the diffuse X-ray background (XRB). 
The knowledge of the fraction of the bolometric luminosity which emerges in the
hard X-ray domain would allow the inclusion of the LIGs in the models of the
XRB. These models are by and large successful in
synthesizing the XRB from the contributions of individual AGNs; they
must however include a dominant contribution from absorbed, type 2 AGNs,
which up to now have been observed directly only at low luminosities and
low redshifts. The zeroth-order extrapolation, which assumes that type
2s and type 1s evolve in the same way (``unified model''), has to face
several discrepancies, and could be cured only by adding extra type 2
sources at intermediate or high redshifts (Gilli et al. 1999). Some of
the LIGs could very well be the required additional sources, but their
mean X-ray properties are still poorly known.

The expected X-ray spectrum, and the relative contribution of the 
X-ray emission to the bolometric luminosity are 
very different in the cases of AGN and starburst dominance,
therefore hard (2-10 keV) X-ray data can be a powerful diagnostic.

The typical X-ray starburst spectrum is a
thermal continuum with temperatures ranging from a fraction of a keV to
several keV (in units of kT), plus several emission lines at low
energies, and a
K$_{\alpha}$ feature of highly ionized
iron at the energy of E$\simeq 6.7$ keV
(Cappi et al. 1999, Persic et al. 1998).
The 2-10 keV luminosity is less than one part in 
10$^3$ of the total emission.

The AGN spectrum between 2 and 10 keV is strongly dependent on the
amount of obscuration suffered by the primary continuum component: if the 
absorbing column density $\rm N_H$ is lower than $10^{24}$ cm$^{-2}$,
the direct
continuum spectrum is visible at energies higher than the photoelectric
cutoff 
E$_C$, whose value depends on N$_H$. The mean direct continuum, as deduced from the
spectra of unabsorbed AGNs, is well represented by a powerlaw with photon
index $\Gamma\sim 1.7$. In addition, a K$_{\alpha}$ iron line is observed
at energies ranging from 6.4 keV (neutral iron) to 6.95 keV
(hydrogen-like iron). The current view is that
the iron line originates both from the accretion disc and from a
reprocessing medium located farther away (1--10 pc):
the first component can be broadened by relativistic effects and is
suppressed if the absorbing column density is higher than several times
10$^{23}$ cm$^{-2}$, the second component is narrow, and its energy
depends
on the ionization degree of the reprocessing medium. 
The equivalent width (EW) of the line is EW $\sim$
100-200 eV with respect to the intrinsic continuum, and can be much
higher (up to some keV) if the continuum is heavily absorbed at the
line energy (while the line, produced in a different region, is still visible).

If the column density is higher than $10^{24}$ cm$^{-2} $
(i.e. the medium is Compton thick), the 
intrinsic emission is completely absorbed and only a reflected
component survives. 
The X-ray spectrum of Compton thick sources is
generally characterized by an Fe line with large
equivalent width ($\sim$ 1 keV) and a reflection dominated continuum 
flatter than the intrinsic spectrum (see Maiolino et al. 1998 for further details).
While the {\it observed}~2--10 keV
contribution to the bolometric luminosity can be higher than 10\% in unabsorbed
AGNs, in Compton thick Seyfert 2s it can be lower than 0.1\%.

Summarizing, the X-ray continuum of a starburst is expected to have a 
thermal high energy cutoff  and a soft spectrum with emission
lines, while an AGN is characterized by a hard power law
with a possible low energy cutoff due to absorption or, in the extremely
absorbed cases (Compton thick), a flat reflection spectrum.
Cold (E$\sim$ 6.4 keV) and broad iron lines, or a large 
relative X-ray luminosity are clear signatures of an AGN, whereas narrow
lines of ionized iron or a relatively weak X-ray emission are possible
in both scenarios. 

In this paper we study the sample of all the luminous infrared
galaxies observed so far in hard X rays (2-10 keV). We 
combine the X-ray information with the IRAS photometry, and
find interesting correlations between X-ray properties and IR
colours. We propose a simple model that matches the observed correlations
and is compatible with the available spectroscopic and spectropolarimetric 
data at near-IR and optical wavelengths.

In Sect. 2 we describe in detail our sample, in Sect. 3 we analyze
the X-ray properties of the sources and the correlation between the X
and IR emission. Finally, in Sect. 4 we discuss our results in the 
framework of a geometrical two-parameter model.
Conclusions and future work are summarized in Sect. 5. In the Appendix
we briefly report the results of our own spectral analysis of those sources
which have not been published elsewhere.
Throughout this paper we use the cosmological parameters H$_0=75$ km
s$^{-1}$ Mpc$^{-1}$
and q$_0=0.5$.

\section{The sample}

The selection of our sample is primarily based on the existence of
data in the
2-10 keV energy range. We collected all the galaxies with
published 2-10 keV data, and those with unpublished observations  in
the ASCA and BeppoSAX archives. Within the X-ray sample we selected all the sources
detected in at least three IRAS bands with L$_{IR} >  10^{11}$ L$_\odot$. 
The final sample consists of 78 objects. The X-ray data were obtained from the literature
for 63 sources, from the ASCA Public Archive for 10, and from BeppoSAX observations 
for the remaining 5. The IR data were obtained from the NASA Extragalactic Database. 

We briefly discuss the 15 unpublished sources in the Appendix.

Table 1 gives all the relevant information about the sample.
The infrared (8-1000 $\mu$m) luminosity has been obtained from the IRAS 
fluxes with the following prescription (see, for instance, Sanders 
\& Mirabel 1996):

\begin{equation}
F_{IR}=1.8~10^{-11}(13.48 f_{12} + 5.16 f_{25} + 2.58 f_{60} +
f_{100}) 
\end{equation}
were $f_{12}, f_{25}, f_{60}, f_{100}$ are the flux densities in the four IRAS
filters expressed in Jansky and $F_{IR}$ is expressed in units of erg
cm$^{-2}$ s$^{-1}$. When we have only an upper limit for one of the four
IRAS points we use one half of the measured upper limit.

\begin{table*}
\centerline{\begin{tabular}{ccccccc|ccccccc}
\hline
name&  z&$\rm L_{IR}^{(a)}$ &
C$_{IR}^{(b)}$  & F$_X^{(c)}$ &
Opt.$^{(d)}$ & X$^{(e)}$&  name&z&  $\rm L_{IR}^{(a)}$ &
C$_{IR}^{(b)}$  & F$_X^{(c)}$& Opt.$^{(d)}$ & X$^{(e)}$\\
\hline
III Zw 2     & 0.0893&   1.5 & 1.52 &   340$^{(19)}$ & 1     & Y & 
MKN 273      & 0.0378&  12.6 & 0.21 &   3.3$^{(1)} $ & 2     & N \\
I Zw 1       & 0.0609&   7.5 & 1.08 &    35$^{(18)}$ & 1     & Y & 
Tol 1351-375 & 0.052 &   1.2 & 1.42 &    38$^{(1)} $ & 2$^*$ & Y \\
I 00198-7926 & 0.0728&  10.3 & 0.74 &    $<1^{(1)} $ & 2$^+$ & N &
MKN 463      & 0.0497&   5.2 & 1.45 &     9$^{(3)} $ & 2$^*$ & Y \\
PHL 0909     & 0.171 &   8.6 & 0.62 &   69$^{(20)} $ & 1     & Y &
PG 1411+442  & 0.089 &   2.2 & 2.00 &  7.6$^{(25)} $ & 1     & Y \\
TON S180     & 0.0620&   1.3 & 1.86 &   45$^{(5)}  $ & 1     & Y &
MKN 1383     & 0.0865&   2.7 & 1.75 &   40$^{(11)} $ & 1     & Y \\
MKN 1048     & 0.0431&   1.9 & 0.75 &  9958$^{(1)} $ & 1     & Y &
MKN 477      & 0.0378&   1.5 & 0.80 &    12$^{(3)} $ & 2$^*$ & Y \\
NGC 1068     & 0.0030&   1.4 & 0.94 &    35$^{(3)} $ & 2$^+$ & Y &
MKN 478      & 0.079 &   2.9 & 0.66 &    25$^{(5)} $ & 1     & Y \\
NGC 1275     & 0.0175&   1.7 & 1.05 &  1140$^{(3)} $ & 2$^*$ & Y &
PG 1444+407  & 0.267 &   7.2 & 1.83 &  5.5$^{(25)} $ & 1     & Y \\
MKN 1073     & 0.0230&   2.3 & 0.35 &  $<$1$^{(1)} $ & 2$^+$ & N &
I 14454-4343 & 0.0386&   4.4 & 0.72 &   2.2$^{(2)} $ & 2$^+$ & N \\
NGC 1365     & 0.0059&   1.6 & 0.31 &    68$^{(9)} $ & 2$^*$ & Y &
MKN 841      & 0.0363&   1.2 & 1.76 &  100$^{(12)} $ & 1     & Y \\
I 03158+4227 & 0.1344&  31.6 & 0.21 &  $<0.7^{(2)} $ & 2     & N &
I 15091-2107 & 0.0446&   2.1 & 0.67 &  260$^{(14)} $ & 1     & Y \\
3C 120       & 0.0330&   1.5 & 1.09 &  455$^{(12)} $ & 1     & Y &
I 15307+3252 & 0.9257& 182   & 0.60 & $<0.7^{(10)} $ & 2     & N \\
I 04154+1755 & 0.056 &   4.3 & 0.37 & $<0.9^{(24)} $ & 2     & N &
ARP 220      & 0.0181&  13.0 & 0.15 &   8.7$^{(7)} $ & 3     & N \\
NGC 1614     & 0.0159&   6.1 & 0.44 &   5.6$^{(1)} $ & 2$^+$ & Y &
PG 1543+489  & 0.4   &  36.8 & 0.76 &    4$^{(5)}  $ & 1     & Y \\
I 05189-2524 & 0.0426&  12.9 & 0.50 &    61$^{(1)} $ & 2$^*$ & Y &
PG 1634+706  & 1.334 & 591   & 0.97 &  7.1$^{(15)} $ & 1     & Y \\
MCG+08-11-11 & 0.0205&   1.2 & 1.30 & 520$^{(11)}  $ & 1     & Y &
NGC 6240     & 0.0245&   6.2 & 0.15 &    19$^{(3)} $ & 2$^+$ & Y \\
H 0557-385   & 0.0344&   1.6 & 4.05 & 140$^{(19)}  $ & 1     & Y &
PG 1700+518  & 0.292 &  23.4 & 0.92 & $<0.8^{(25)} $ & 1     & Y \\
I 07598+6508 & 0.1488&  26.7 & 0.63 &  $<0.8^{(1)} $ & 2$^+$ & N &
I 17020+4544 & 0.0604&   2.9 & 0.75 &    80$^{(5)} $ & 2$^*$ & Y \\
PG 0804+761  & 0.10  &   3.5 & 2.8  &   88$^{(25)} $ & 1     & Y &
I 17208-0014 & 0.0428&  21.6 & 0.11 &     3$^{(1)} $ & 3     & N \\
NGC 2623     & 0.0185&   3.1 & 0.16 &  $<0.8^{(1)} $ & 2     & N &
MKN 507      & 0.0559&   1.0 & 0.37 &     5$^{(3)} $ & 2$^*$ & Y \\
PG 0844+349  & 0.064 &   1.2 & 2.50 & 23.6$^{(25)} $ & 1     & Y &
NGC 6552     & 0.0262&   1.2 & 0.83 &     8$^{(3)} $ & 2$^+$ & Y \\
I 08572+3915 & 0.0582&  12.4 & 0.48 &  $<0.7^{(1)} $ & 2$^+$ & N &
I 18325-5926 & 0.0202&   1.0 & 0.87 &    84$^{(3)} $ & 2$^*$ & Y \\
I 09104+4109 & 0.4420& 107   & 1.26 &     5$^{(4)} $ & 2$^+$ & Y &
I 18293-3413 & 0.0182&   5.3 & 0.22 &   0.9$^{(1)} $ & 3     & N \\
NGC 3256     & 0.0091&   3.6 & 0.36 &     7$^{(8)} $ & 3     & N &
3C 390       & 0.057 &   2.1 & 1.45 &  170$^{(12)} $ & 1     & Y \\
MCG+12-10-067& 0.0328&   1.0 & 0.5  &     30$^{(3)}$ & 2$^*$ & Y &
H 1846-786   & 0.0743&   1.9 & 0.97 &  170$^{(19)} $ & 1     & Y \\
I 11058-1131 & 0.0548&   1.8 & 0.83 &   3.9$^{(2)} $ & 2$^+$ & Y &
I 19254-7245 & 0.0617&  10.5 & 0.44 &  1.9$^{(24)} $ & 2$^+$ & Y \\
PG 1114+445  & 0.144 &   4.5 & 1.58 &   21$^{(25)} $ & 1     & Y &
I 20210+1121 & 0.0564&   6.7 & 0.83 &   2.9$^{(2)} $ & 2$^+$ & Y \\
NGC 3690     & 0.0104&   6.1 & 0.20 &   15$^{(17)} $ & 3     & N &
MKN 509      & 0.0353&   1.6 & 1.07 &  440$^{(12)} $ & 1     & Y \\
PG 1148+549  & 0.9690&  63.1 & 1.20 &   27$^{(25)} $ & 1     & Y &
I 20460+1925 & 0.1810&  28.6 & 1.20 &    15$^{(3)} $ & 2$^*$ & Y \\
WAS 49b      & 0.0637&   1.3 & 1.59 &     12$^{(1)}$ & 2$^*$ & Y &
I 20551-4250 & 0.0428&   9.7 & 0.30 &   3.5$^{(6)} $ & 3     & N \\
PG 1211+143  & 0.0809&   3.5 & 2.32 & 27.8$^{(25)} $ & 1     & Y &
II Zw 136    & 0.0617&   1.8 & 1.73 &   53$^{(11)} $ & 1     & Y \\
3C 273       & 0.1583&  49.9 & 0.87 &  1500$^{(21)}$ & 1     & Y &
IC 5135      & 0.0160&   2.1 & 0.26 &    5$^{(23)} $ & 2$^+$ & Y \\
MKN 231      & 0.0422&  32.4 & 0.50 &     6$^{(7)} $ & 1     & Y &
I 22017+0319 & 0.0611&   4.1 & 1.24 &    36$^{(3)} $ & 2$^*$ & Y \\
MCG-03-34-064& 0.0165&   1.3 & 0.97 &   47$^{(22)} $ & 1     & Y &
NGC 7212     & 0.0266&   1.2 & 0.53 &    9.8$^{(1)}$ & 2$^*$ & Y \\
I 13224-3809 & 0.0667&   3.4 & 0.33 &   5.2$^{(5)} $ & 1     & Y &
3C 445       & 0.0562&   1.8 & 1.68 &   77$^{(13)} $ & 1     & Y \\
 NGC 5135    & 0.0130&   1.5 & 0.31 &     2$^{(3)} $ & 2$^+$ & Y &
NGC 7469     & 0.0170&   4.0 & 0.41 &  290$^{(12)} $ & 1     & Y \\
I 13349+2438 & 0.1076&  16.6 & 2.75 &   57$^{(16)} $ & 1     & Y &
I 23060+0505 & 0.173 &  25.7 & 0.75 &    15$^{(3)} $ & 2$^*$ & Y \\
I 13451+1232 & 0.1210&   5.1 & 0.70 &    10$^{(24)}$ & 2$^*$ & Y &
I 23128-5919 & 0.0446&   9.4 & 0.30 &   3.5$^{(6)} $ & 2$^+$ & N \\
MKN 266      & 0.0279&   2.8 & 0.31 &   5.3$^{(1)} $ & 2$^+$ & Y &
NGC 7674     & 0.029 &   3.2 & 0.70 &     5$^{(3)} $ & 2$^+$ & Y \\ 
\hline
\end{tabular}}
\caption{\footnotesize{$^{(a)}$ Luminosity in units of 10$^{11}$ L$_{\odot}$;
$^{(b)}$ Infrared colour, defined as C=$\frac{2 f_{25}}{f_{60}}$. $^{(c)}$ 
2-10 keV flux in of units 10$^{-13}$ erg
cm$^{-2}$ s$^{-1}$; $^{(d)}$ Optical classification (1: 
Quasars and Seyfert 1s, 2: Seyfert 2s and LINERs, 3: Starbursts. Type 2
objects with an asterisc and a cross are Compton thin and thick,
respectively,
according to their X-ray
spectral properties; the others are unclassified in the X rays.
$^{(e)}$ X-ray evidence for AGN (Y=Yes; N=No).
References: $^1$ This work;
$^2$ Ueno et al. 2000; $^3$ Bassani et
al. 1999; $^4$ Franceschini et al. 2000; $^5$ Vaughan et al. 1999; $^6$ Misaki et
al. 1999; $^7$ Iwasawa 1999; $^{8}$ Moran et al. 1999;
$^9$ Risaliti et al.
2000; $^{10}$ Ogasaka et al. 1997; $^{11}$ Lawson \& Turner 1997;
$^{12}$ Reynolds 1997; $^{13}$ Sambruna et al. 1998; $^{14}$ Awaki et al. 1991;
$^{15}$ Nandra 1995; $^{16}$ Brandt et al. 1997; $^{17}$ Zezas et al.
2000; $^{18}$ Leighly 2000; $^{19}$ Turner \& Pounds 1989; $^{20}$ Ghosh \&
Soundararajaperumal 1992; $^{21}$ Cappi et al. 1998; $^{22}$ Ueno 1997;
$^{23}$ Risaliti et al. 1999; $^{24}$ Imanishi \& Ueno 1999; 
$^{25}$ George et al. 2000.}}
\end{table*}

The sample is very heterogeneous, and to assess the validity of any
conclusion one must consider the selection criteria of the many 
subsamples which were merged.

We note from Table 1 that most objects are optically classified as AGNs 
and some of them are among the best known and widely studied type 1 and 
type 2 AGNs. In fact, 46 objects out of 78 were selected by the original
authors because of their well known AGN activity, irrespective of their 
IR properties. The remaining 32 were instead selected for observation
because of their high IR 
luminosity (many of them are ULIGs, i.e. sources with $\rm L_{IR} > 10^{12}$ 
L$_{\odot}$, see Fig.~1). Nevertheless, also this second half is 
strongly biased toward AGN-dominated sources: many targets were selected 
because of the presence of some AGN indicator, such as broad lines in the 
polarized optical spectrum, or compact radio cores with high brightness 
temperature. A few objects had no AGN indicator of any sort, and were 
observed to assess the energy source in a ``generic'' LIG; they were chosen
among bright IR sources without previous hard X-ray data, so they are 
underluminous in X~rays with respect to the average LIG, and in Fig.~1 
they are adiacent to well known starbursts.

\begin{figure}\centerline{\resizebox{\hsize}{!}
{\includegraphics{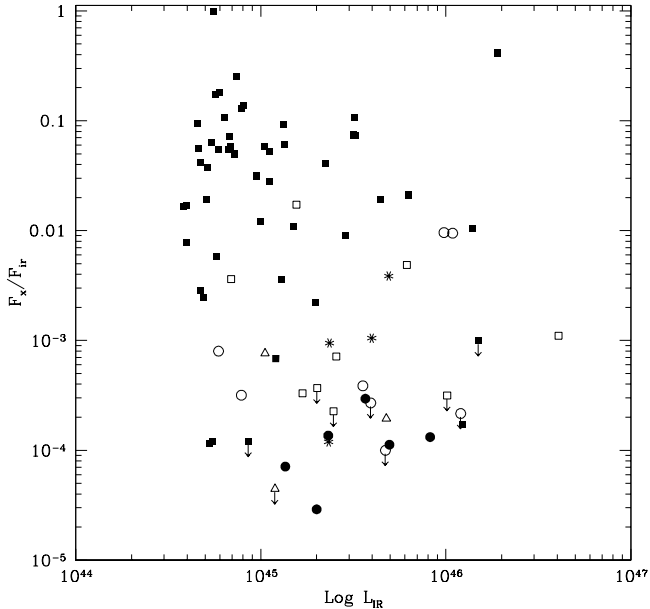}}}
\caption{\footnotesize{The 2-10 keV to IR flux ratio plotted as a
function of the IR luminosity.
Graphical codes represent the different selection criteria: filled
squares: known AGNs; open squares: high IR luminosity and broad polarized lines
in the optical; open triangles: high IR luminosity and radio emission with
brightness temperature T$_B > 10^7$ K; stars: mergers; open circles: only high
IR luminosity; filled circles: known starbursts.}}
\end{figure}

In Fig. 1 we show the X-ray to IR flux ratio versus the infrared luminosity.
The different codes correspond to the selection criteria which were the
drivers of the X-ray observation.
We note that the much discussed correlation between the IR luminosity and 
the occurrence of AGN activity (Sanders \& Mirabel 1996,
 Genzel et al. 1998, Lutz et al. 1998)
does not appear in Fig.~1, just because of the strong bias in favor
of AGN-dominated sources discussed above.
Well known AGNs and well known starbursts\footnote{ The classification of
ARP~220 is ambiguous (see Smith et al.
1998, Genzel et al. 1998, Kim, Veilleux \& Sanders 1998); here we have adopted
the starburst classification, but moving the source to the Seyfert 2 class would
not change any of our conclusions.}
cluster very clearly
in the upper and lower half of the diagram, respectively,
 confirming the diagnostic
value of the X-ray to IR luminosity ratio. We further note that all
flavours of sources are represented in the sample, although their relative
numbers are altered with respect to a proper, unbiased
 sample, and they are spread
over a large region of the parameter space: thus the true distributions
of the various observables among LIGs
 cannot be derived from our data set, but any
correlation among these observables is likely to be real, as discussed in 
the following.

Finally, we note that some of the AGNs in the sample are
radio loud. However, with the exception of a blazar (3C273),
their radio ``loudness'' does not appear to introduce
any anomalous behaviour, in terms of their X-ray or IR properties,
with respect to the radio quiet sources (which are the
great majority); this will be discussed in Sect.~3.2.

\section{X and IR properties of the sample}

\subsection{X~rays}

{\it 1) X-ray absorption in AGNs}\hfill\break
The X-ray properties of our sample are extremely varied. Most of the objects
optically classified as type 1 Seyferts of QSOs are bright X-ray emitters
(relative to their IR luminosity), and have no signs of large absorption in
their spectra (with the exception of the three
 Broad Absorption Line QSOs, which will be
discussed later). A significant fraction of the narrow line AGNs are also
relatively bright in the X rays and have spectra characterized by
Compton thin absorbtion 
($\rm N_H < 10^{24}~cm^{-2}$). In other narrow line AGNs the 2--10 keV spectrum
is very flat and/or the Fe line at 6.4 keV has a
large equivalent width ($\sim$ 1 keV); as discussed in the Introduction both
these features identify reflection dominated AGNs whose direct X-ray radiation
is obscured along our line of sight by Compton thick gas
($\rm N_H > 10^{24}~cm^{-2}$). For 11 narrow line AGNs, most of which are new
sources whose analysis is presented in the Appendix, 
there is no spectral information in the hard X-rays, and 
only a weak X-ray detection or an upper limit are available.
In these cases either the AGN is heavily
obscured (Compton thick), or it is intrinsically weak (and the IR
luminosity dominated by the starburst), or a combination of the two.
If the flux of the (reddening corrected) [OIII]$\lambda$5007
narrow line is assumed to be a fair, isotropic indicator of the
intrinsic luminosity, then the ratio between the flux of this line and the
observed X-ray flux provides information about the absorption affecting
the nuclear X-ray source. In particular, as discussed in Bassani et al.
 (1999) and Maiolino et al. (1998),
a flux ratio X/[OIII] lower than unity is typical of
Compton thick sources.
Of the 11 X-ray faint AGNs, 7 have measured [OIII] and Balmer decrement; their
X/[OIII] ratio is always $<1$, and indicates that these must be Compton thick,
reflection dominated sources. For the remaining 4 AGNs, no conclusion can be
drawn on the X-ray absorption and luminosity.


\medskip
\noindent {\it 2) X-ray spectral indices:}\hfill\break
We have studied the (instrinsic) X-ray spectral indices
of a sub-sample of objects with
an X/IR flux ratio higher than 10$^{-2}$. As discussed in the following,
this condition selects mostly broad line AGNs with no X-ray absorption
or narrow line AGNs with mild X-ray absorption, so that the intrinsic
spectral index can be reliably derived by correcting for the observed
photoelectric cutoff. Also, we only considered sources with X-ray observations
down to 0.5 keV, in order to detect possible soft thermal components, and
dropped all the Narrow Line Seyfert 1s, which are known to have an intrinsic
spectrum different (steeper) from ``normal'' Seyferts.
This sub-sample was divided
in two groups according to the IR colour\footnote{In this work
we use an infrared
colour defined as the ratio between the {\it fluxes} in the IRAS 25 $\mu$m and
60$\mu$m bands, that, according to Eq. 1, is given by C$=\frac{2
f_{25}}{f_{60}}$.}, C, being larger or smaller than 1.1. 
As shown in Fig. 2, objects with 
cold IR colour have an X-ray spectrum softer than usually measured in
Seyfert~1s and quasars: the photon index of 6 sources out of 8 
is $1.98 < \Gamma <2.5$. 
As for the group with warm IR colour, only 2 
sources out of 14 have a photon index $\Gamma >2$, 
while the remaining  12
have $\Gamma < 2$ (Fig. 2).
\begin{figure}
\centerline{\resizebox{\hsize}{!}
{\includegraphics{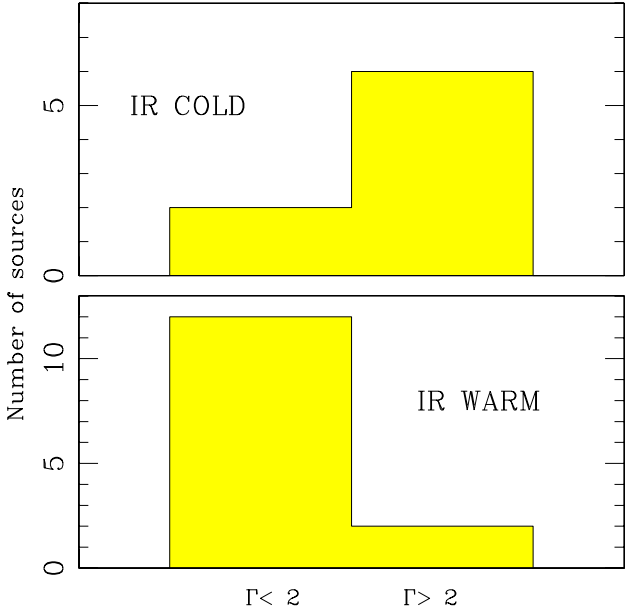}}}
\caption{\footnotesize{Photon index distribution of the sources of our
sample having F$_X$ / F$_{IR} > 0.01$. Upper panel, IR ``cold'' sources;
lower panel, IR ``warm'' sources.}}
\end{figure}
We must caution that the data are not homogeneous: for many objects 
we have the result of the spectral fitting with a single powerlaw,
while for others the fit includes a separate thermal component to account 
for the soft excess. In the latter cases we re-fitted the composite model 
spectra with a single powerlaw, to make the results comparable with each
other. Of course our simple procedure is far from precise, and a more
detailed analysis would require a new fit to the original data. 

This finding on the spectral indices
is relevant to the explanation we propose for the 
correlations of Fig.~3, and will be discussed in the next Section.
\medskip

\noindent {\it 3) The BeppoSAX subsample:}\hfill\break
In the Appendix we report the analysis of 5 BeppoSAX observations of LIGs. 
Three of them are ULIGs with compact radio emission and
brightness temperature T$_B>10^7$ K, one (NGC 1073) is a 
Seyfert 2 with a strong [O III] emission; the last one (IRAS 00198-7926) is
a Seyfert 2 with warm IR colour. In addition to these,
our sample contains 8 more type 2 sources observed with BeppoSAX:
IRAS 11058-1131, IRAS 22017+0319, IRAS 20210+1121, IRAS 14454-4343, studied
by Ueno et al. (2000); NGC 1365 (Risaliti et al. 2000),
IRAS 09104+4109 (Franceschini et al. 2000), NGC 1068 (Matt et al. 1997)
and NGC 6240 (Vignati et al. 1999). The 13 sources listed 
above constitute a first small sample 
observed up to 200 keV by means of the PDS instrument.
Among these sources only three, IRAS 22017+0319, NGC 1365 and NGC 6240, clearly 
show a
direct emission, with a typical Seyfert 2 spectrum that extends
from 1 to 200 keV
and an absorbing column density of 4$\times 10^{22}$ cm$^{-2}$, 
4$\times10^{23}$ cm$^{-2}$ and 2$\times10^{24}$ cm$^{-2}$,
respectively. The remaining 10 sources are 
very weak in X~rays (with the exception of NGC 1068, that has a good
signal-to-noise spectrum, even if it is completely Compton-thick up to
200 keV, Matt et al. 1997), and have a flat 2-10 keV spectral index (when at 
all constrained), all hints of heavy obscuration. For 6 out of these 10
sources we have a detection in the PDS (marginal in all objects but NGC
1068) that is consistent with the extrapolation towards higher energies
of the 2-10 keV best
fit. The most straightforward interpretation of these objects is that
they are Compton thick and reflection--dominated even in the 10--200 keV band,
which implies 
a column density $> 10^{25}$ cm$^{-2}$. Alternatively, some of
these AGNs might be fading and their cold reflection dominated spectrum
could be the echo of their past activity, as observed in NGC 2992
(Weaver et al. 1996, Gilli et al. 2000), Mkn 3 (Iwasawa et al. 1994) and
NGC 4051 (Guainazzi et al. 1998). In particular, the latter might be the
case for Mkn 273.
For one object, IRAS 09104+4109, a marginal
detection of an excess was obtained, implying $\rm N_H \sim 5\times 10^{24}$
cm$^{-2}$. For the remaining three sources we can only put a lower limit
to $\rm N_H$ of $10^{24}$ cm$^{-2}$, either because the PDS upper limit is
inconclusive, or because there is a confusing source in the PDS field of
view.  

We will discuss in Sect. 4.1 the implication of such a large number of 
reflection dominated sources.

\subsection{X-IR colour-magnitude diagram}

An infrared indicator of the presence of an AGN is the ratio between the
fluxes at 25 and 60 $\mu$m (see Note 1 for its definition).
The 60 $\mu$m emission is mainly due to
reprocessing of the UV-optical radiation by dust heated at
intermediate temperatures ($\sim$50K), 
that is present both in starbursts and in the circumnuclear torus of AGNs. 
The 25 $\mu$m emission is due to warmer dust (T$\sim$ 100 K) that is more
abundant in the central regions of AGNs.  
We note that this simple scheme is altered by the obscuration 
thought to be present in type 2 AGNs, since the 25 $\mu$m emission can 
be absorbed as well (assuming a Galactic dust to gas ratio, the optical 
depth at 25 $\mu$m is $\tau > 1$ for $\rm N_H > 10^{23}$ cm$^{-2}$,
Draine 1989). This effect 
will be analyzed in detail in the following Section.

In Fig. 3 we plot the X/IR flux ratio versus the 25-60 $\mu$m 
colour\footnote{We excluded from the plot in Fig. 3 four of the objects 
listed in Table~1: 3C~273, whose infrared emission is known to be dominated 
by a blazar component, at least in outburst; and IRAS 15307+3252, PG 1148+549 and
PG 1634+706, that are the only three objects with redshift as high as 
$\sim $1, where the K-correction could be important.}. The
different symbols are related to the optical classification (Seyfert 1/QSO,
Seyfert 2, starburst) and to the Compton thin / Compton thick classification
derived from the X-ray analysis discussed in Sect.~3.1.

A clear correlation is apparent in Fig. 3: type 1 AGNs have preferentially 
high X/IR ratios and warm infrared colours. 
Moving towards lower 25/60$\mu$m ratios we find lower X/IR ratios and an 
increasing fraction of obscured AGNs at first, and of starbursts afterwards.
Together with the correlation we note also a large scatter, which will be
discussed in detail in the next Section.

\begin{figure}\centerline{\resizebox{\hsize}{!}
{\includegraphics{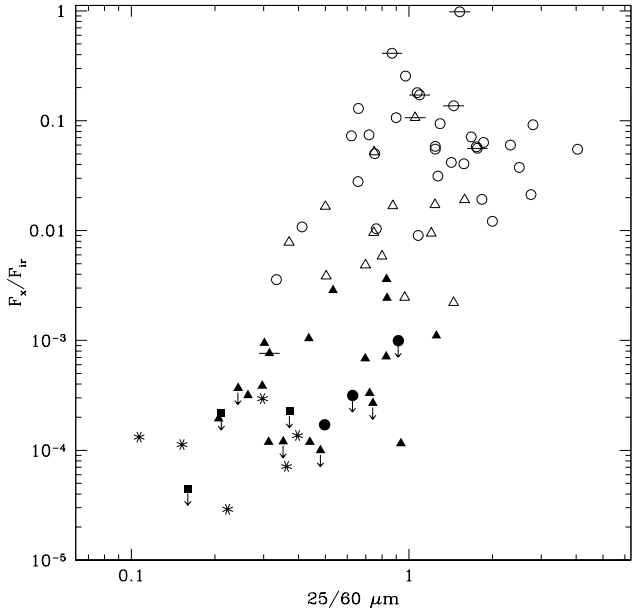}}}
\caption{\footnotesize{The X/IR flux ratio of LIGs versus the 25-60 $\mu$m 
colour $\frac{2\times f_{25}}{f_{60}}$, with the same notations 
as in Table~1. Symbols: open circles: Seyfert 1s; open triangles: 
Compton thin Seyfert 2s; filled triangles:
Compton thick Seyfert 2s; filled circles: broad absorption line quasars;
stars: starbursts; filled squares: Sy2s with no information on the relative
AGN/starburst contribution (see text). Radio--loud sources are marked with a
horizontal segment.}}
\end{figure}

The result presented in Fig.~3 is in qualitative agreement with general
expectations. In addition, it poses interesting quantitative constraints
on the physics of LIGs. Since the boundaries depicted in Fig.~3 are 
crucial for our subsequent analysis, we must: (1) check if the lack of 
objects with low X-ray flux and warm IR colour and/or with high 
X-ray flux and cold IR colour can be ascribed to the 
selection biases illustrated in the previous Section; (2) compare
the LIGs with non-LIG sources, in order to see if the 
correlation of Fig. 3 is a characteristic of the former or is
present in the generic hard X-ray extragalactic population.

(1) The shortage of objects in the top-left region of 
Fig. 3 is very likely real, since
X--ray loud objects should have been detected easily.
For instance, the same region is populated in Fig. 4.
Concerning the lack of sources in the bottom-right part 
of the Figure, we have studied a sample of IR-warm LIGs not observed in 
X~rays: from the IR colour and the IR flux we have estimated the minimum 
X-ray flux that a source should have in order to follow the correlation of 
Fig. 3. If the minimum flux were higher than -say-
$\sim 10^{-11}$ erg cm$^{-2}$ s$^{-1}$ for a significant number of sources,
then we should
conclude that probably these objects do not match the correlation, because 
it would be very improbable that such X-ray loud sources escaped detection
in all previous X-ray surveys. 
We used to this purpose the catalog of de 
Grijp (1985), which is a selection of the IR-warmest sources in the 
IRAS Point Source Catalog. We found that for all the LIGs of this
catalog not included in our sample the minimum required X-ray flux 
is lower than about 10$^{-12}$ erg cm$^{-2}$ s$^{-1}$. Summarizing, there are
no obvious bias effects which could ascribe
the observed correlation to selection problems.

(2) The second point has been investigated by checking if a similar correlation
exists also for sources with L $< 10^{11}$L$_\odot$. Our control sample of 
non-LIGs is composed of $\sim 50$ AGNs with hard X-ray data available in the 
literature. The sample is not complete neither representative of a
particular class of sources, but it shows nonetheless (see Fig. 4) that the
absence of strong X-ray emitters with cold IR colours is a feature typical of
the LIGs. On the contrary, the lack of IR warm and X-ray weak objects could be a
general property of hard X-ray sources.

\begin{figure}\centerline{\resizebox{\hsize}{!}
{\includegraphics{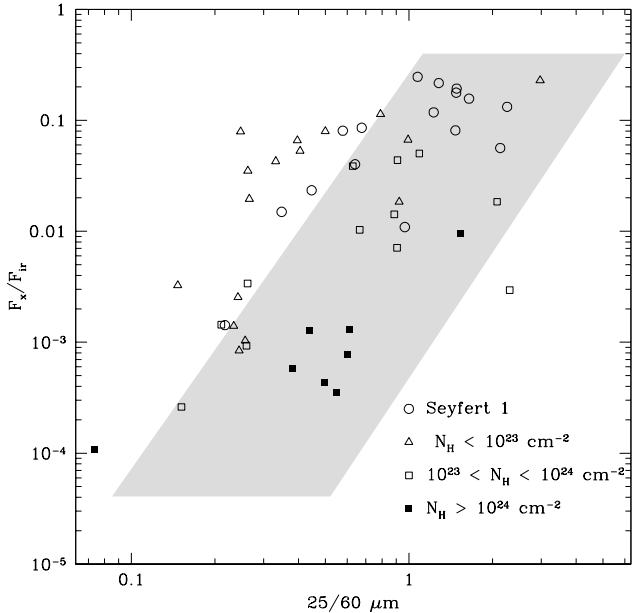}}}
\caption{\footnotesize{The same as Fig.~3 for a sample
of galaxies with $\rm L_{IR} < 10^{11} L_{\odot}$
The shaded region is the one occupied by the LIGs of Fig.~3.}}
\end{figure}

Finally, we note that the radio loud (broad line) AGNs, marked with a horizontal
segment
in Fig. 3, do not have an anomalous behaviour with respect to type 1
radio quiet objects. Instead, the three broad absorption line (BAL)
QSOs (identified
with a filled circle in Fig. 3) are characterized by a low X-ray luminosity
compared to the other broad line AGNs. This is a well known property of
BAL QSOs and it is ascribed to absorption associated to the outflowing
medium responsible for the BALs (Brandt et al. 1999,
Crenshaw et al. 1999). 

\section{Discussion, and a basic scenario}

Our approach is to compare the observed distribution of Fig.~3 with
various combinations of starburst and AGN spectra, the latter absorbed 
according to different prescriptions.

The horizontal line in the upper part of Fig.~5 represents a model with 
only the AGN component: the rightmost point, at coordinates 2.1 (colour) 
and 0.1 (flux ratio), is a typical location for bright Seyfert~1s.
This ``starting point'' cannot be the average of the Seyfert~1s in our
sample, because some of them are contaminated by starburst activity. On 
the other hand, mean values of lower luminosity objects could contain a
contribution from the host. We then adopted as a template the bright
Seyfert~1 IC4329A, that is not a LIG, but has an infrared luminosity 
($\rm L_{IR}\simeq 6\times$ 10$^{10}$L$_{\odot}$) significantly higher than normal 
galactic values, no indication of starburst activity and a well studied,
warm IR spectrum.
The X-ray source is modelled as a powerlaw with photon index $\Gamma=1.7$,
and the absorbing material --covering the X-ray source in the same way as 
the infrared source-- is assumed to have a Galactic dust--to--gas ratio 
and extinction curve. In the infrared, only the 25$\mu$m flux is affected
by the absorption, while the 60$\mu$m flux and the bolometric flux remain
constant.

\begin{figure*}
\centerline{\resizebox{\hsize}{!}
{\includegraphics{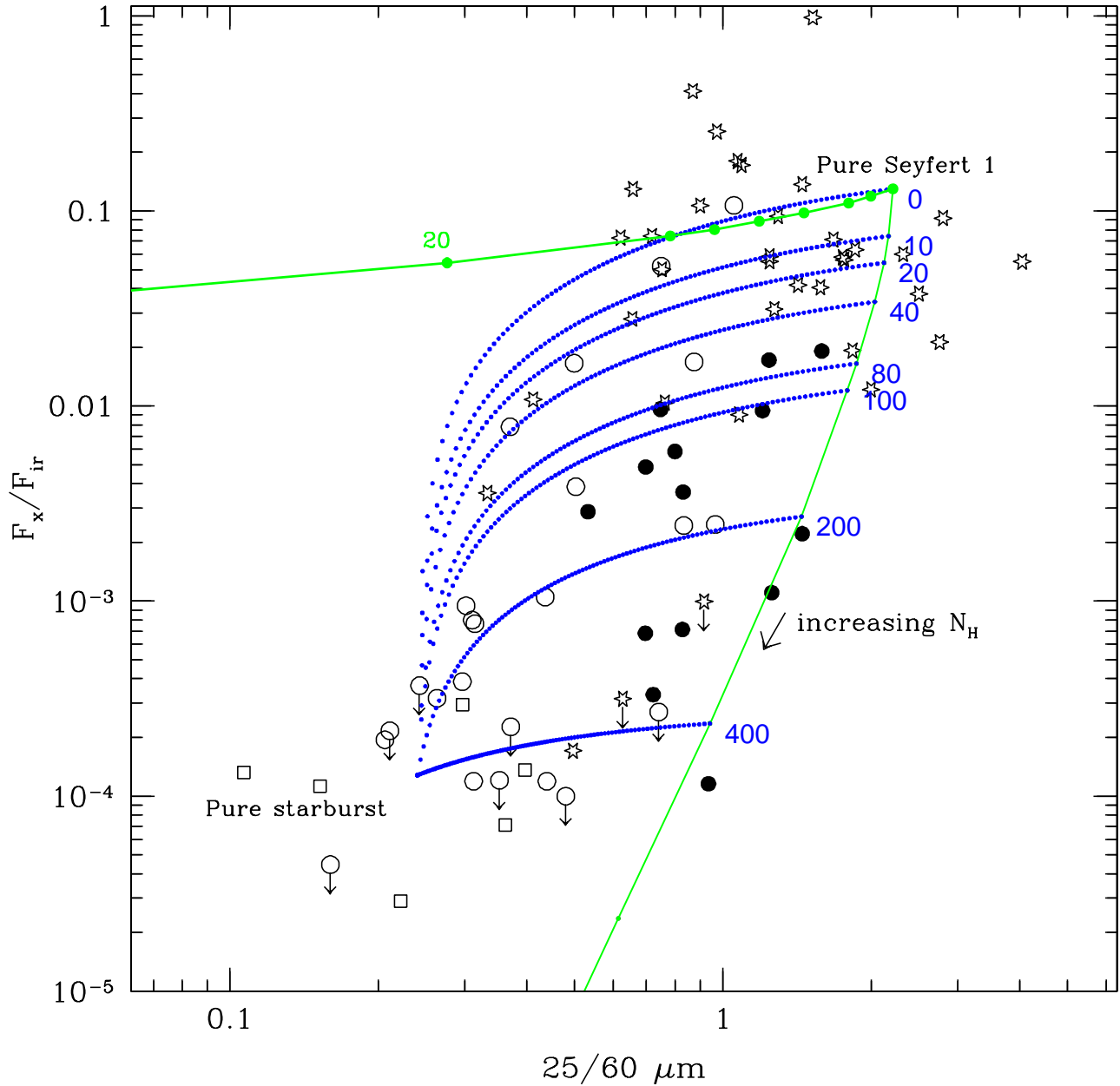}}}
\caption{\footnotesize{The results of the double absorber model: the
gray line
on the right represents AGN dominated sources, with absorbing X-ray
column
density (in units of 10$^{22}$ cm$^{-2}$) increasing as indicated. The
dotted curves are mixing curves of AGN and starburst emission, with the
latter
increasing towards the left. Filled circles are the sources in our
sample with observational evidence of broad lines in the polarized
optical or near infrared spectra. The gray line at the top represents a
model where the same material absorbs both the IR and the X-ray emission
of an AGN, with a Galactic dust--to--gas ratio (see text for details).}}
\end{figure*}

The model is obviously incorrect,  and does not fit the observed trend
even for those many objects that are known to
be dominated by the AGN component. The prediction is a fast decrease
of the colour parameter at an almost constant X/IR flux ratio, whereas
the data show that the two quantities decline together. To decrease
the X-ray flux by an amount comparable to the 25$\mu$m flux,
the absorbing material must have properties different from the Galactic gas, or 
it must be distributed differently against the X-ray and the IR sources.

The warm dust emitting the bulk of the 25 $\mu$m emission is located at
several pc from the nuclear source and therefore suffers 
a much lower absorption than the nuclear X-ray
source. This might explain the discrepancy between model and data.
In a scenario like that of Fig.~6, if the medium-IR source is at a fixed 
height above the plane, one could crudely assume that it sees a
fixed small fraction of the absorption seen by the X-ray source, for a
substantial range of inclination angles.

\begin{equation}
{\rm N}_H{\rm(IR) = {\it k}\times N}_H\rm(X)
\end{equation}
where {\it k} is a constant. We have adopted a Galactic gas--to--dust ratio 
and changed the IR absorption with respect to the X-ray one, but we
could have taken an identical $\rm N_H$ toward both sources and a lower
$A_V/{\rm N}_H$. Then the constant {\it k} in the equation above, which
in our model is a geometrical factor, would represent the correction 
to the dust--to--gas ratio.

The oblique rightmost line in Fig.~5 represents the double absorber,
AGN dominated model described in Fig.~6. The N$_{H}$ values indicated
by the labels refer to the X-ray absorption, while the parameter {\it k}
was adjusted ({\it k} $\sim 70$) 
to fit the boundary of the populated region. In order to
add a starburst component, we took the average colour and X/IR flux
ratio of the six starburst-dominated sources in our sample, since in 
these cases the possible contamination from additional components (such 
as a very obscured AGN) should be negligible. The dotted
curves give different degrees of mixing,
starting from the right with objects 100\% dominated by a more or less
absorbed AGN.

In summary, the position of an object along the main extension ($\sim$
vertical) of 
the plot is a measure of the amount of absorption incurred by the AGN
component, while the position across the plot, along the horizontal
direction, is a measure of the amount of mixing with a starburst 
component. 

This interpretation is supported by the optical identifications. However,
there are some further points which support the proposed picture:\hfill\break
1) According to our scheme, the type 2 sources having the warmest IR colour 
allowed by their X/IR flux ratio are dominated by the AGN component, with 
low absorption on large scales. Therefore we expect to find broad
polarized or near-IR broad
lines more often in these objects than in colder ones (see also Heisler,
Lumsden \& Bailey 1997).
In objects with colder IR colors the starburst is more and more important, and the AGN lines 
could be diluted by the radiation contributed by the starburst, or suppressed 
by the large scale absorption associated with it. 
There are only a few type 2 sources with the required data
(spectropolarimetry and/or near-IR spectra), but those
with broad polarized or near-IR broad
lines all lie along the AGN locus in Fig.~5, as expected.\hfill\break
\begin{figure}[h]
\centerline{\resizebox{\hsize}{!}
{\includegraphics{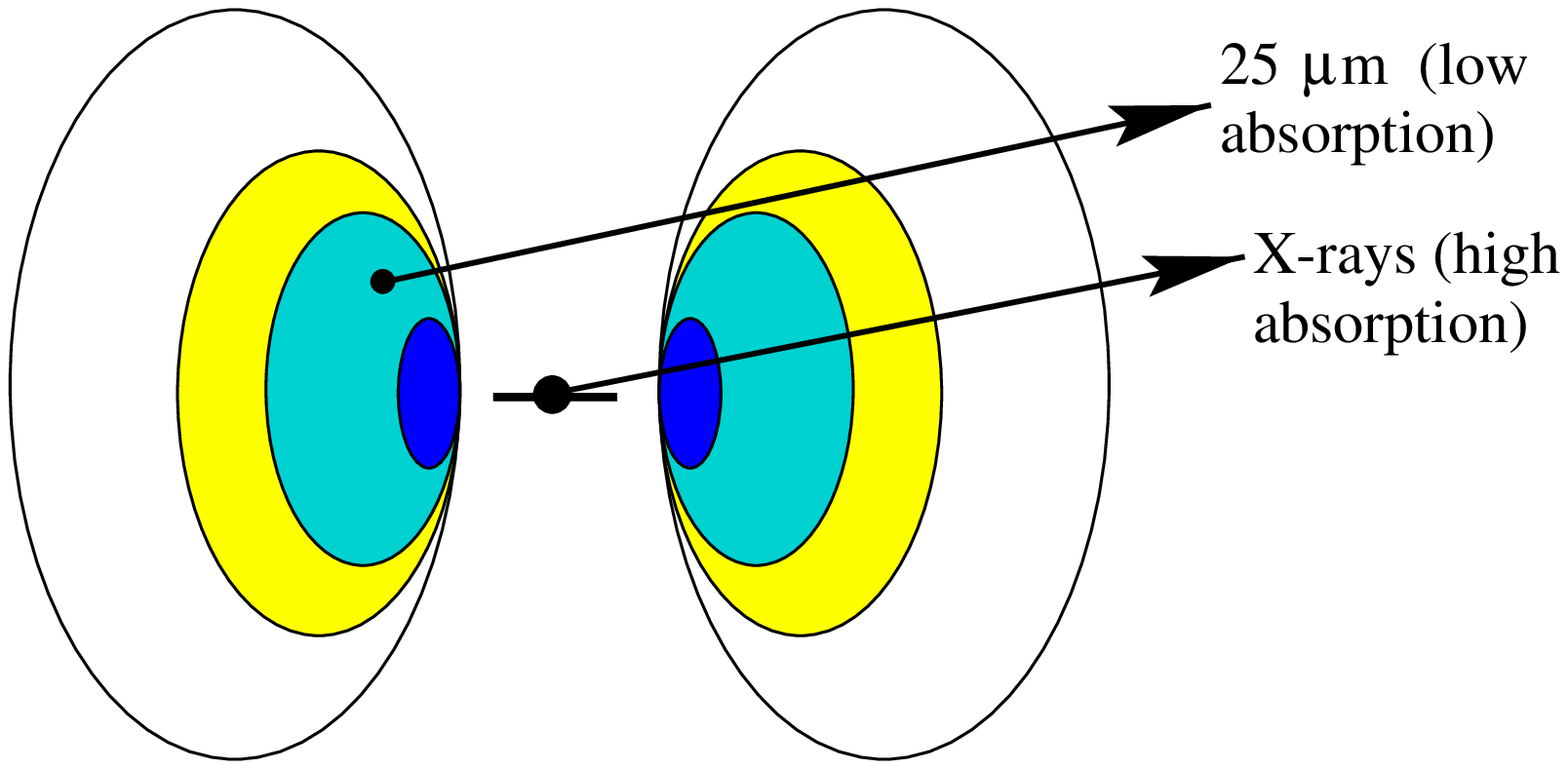}}}
\caption{\footnotesize{A schematic view of the central region of an AGN:
the X~rays from the nucleus are heavily absorbed, while the 25$\mu$m
emission, originating from the warm dust of the torus,
is much less obscured.}}
\end{figure}
2) The X-ray signature of the starburst contribution is a thermal component 
with typical kT values of 0.1-3 keV. Therefore, in this case,
 the overall X-ray spectrum should be 
steeper, and the fit with a single powerlaw should give a photon
index larger than in typical AGNs, say $\Gamma > 2$. The evidence presented 
in Fig.~2 goes exactly in this direction: the large majority of the IR-cold
sources have steep (starburst-like) X-ray indices while
IR-warm sources have flatter (AGN-like) indices.

Our model seems to provide a qualitative explanation for several observed
trends. However, the Sy1s located above and to the right of the ``starting
point'' of Fig. 5 give an idea of the intrinsic spread of these two quantities;
the spread is comparable to the horizontal width of the plot, and indicates that
care must be exercised in drawing strong conclusions from the IR colour.

\subsection{Non-detections at E $>$ 20 keV}

In Sect. 3.1 we briefly presented a subsample of 13 LIGs with
observations up to 200 keV, provided by the PDS onboard BeppoSAX.
Observations at several tens of keV allow the exploration of the 
$\rm N_H$ range 10$^{24}$ cm$^{-2} < \rm N_H < 10^{25}$ cm$^{-2}$, since
these column densities suppress completely the direct radiation up
to about 10 keV, but become translucent at higher energies. Well known 
examples are NGC 4945 (Done et al. 1996) and the Circinus 
galaxy (Matt et al. 1999). A significant lack of objects with 10$^{24}$ 
cm$^{-2} < \rm N_H < 10^{25}$ cm$^{-2}$ was already pointed out in a sample 
of optically selected Seyfert 2s (Risaliti et al. 1999). 
Here we seem to find the same result for the LIGs: among all sources obscured 
between 2 and 10 keV, and having PDS observations, 6 are reflection
dominated up to
100 keV, implying an $\rm N_H > 10^{25}$ cm$^{-2}$ (or, alternatively, a
fading nucleus); only one has 10$^{24}$
cm$^{-2} < \rm N_H < 10^{25}$ cm$^{-2}$, while for the others no
firm conclusion can be drawn.

This finding, if confirmed for a larger and well selected sample, would
have important consequences in the synthesis models of the X-ray
background. If a large fraction of sources are reflection dominated up
to 100 keV,
the integrated flux of the absorbed AGNs is
dimmed and they are on average less detectable at a given flux. 
Therefore, the ratio
between absorbed and unabsorbed AGNs should be increased with respect
to the local value to account for the hard XRB intensity and
hard X-ray source counts (Gilli et al. 1999). 

\subsection{An unbiased sample of LIGs}

As we already discussed in Sect. 2, the sample studied in this paper
is heterogeneous and incomplete.
However, since it contains all the X-ray data currently
available, we can extract from it the least biased subsample that it is
possible to assemble at present.
To do this, we ordered the LIGs of the Bright Galaxy Sample 
(BGS, Soifer et al. 1987, Sanders et al. 1995) according to their {\it
total} infrared flux, and selected the first $n$ of them so as to
maximize the fraction of objects with hard X-ray data.
It is worth noting that by selecting in total IR flux we remove a
bias that affects the BGS catalogue (and any randomly selected
subsample). In fact the BGS is flux-limited at 60$\mu$m and
therefore the effective cut in terms of total IR flux is higher for the AGNs
(which on average have a lower fraction of the IR luminosity emitted at
60$\mu$m) than for the starbursts (which have the peak of their
emission at 60 $\mu$m). The magnitude of the effect can be
estimated by using the sources listed in Table~1 and the equation quoted
in Sect. 2: we find that the fraction of the IR flux emitted at 
60$\mu$m is 50-55\% for the starbursts and 15-20\% for the IR-warmest 
AGNs. So, in order to eliminate the bias, one has to remain a factor
of at least $\sim 3$ above the limit of the parent catalog.

The number of objects included in the subsample
 is 23, which corresponds to a limiting total IR
flux of 1.6 10$^{-9}$ erg cm$^{-2}$ s$^{-1}$. In this way the
fraction of X-ray observed sources is highest (15 out of
23, i.e. 65\%) and the lowest value of the 60$\mu$m flux density is
15.5 Jy, indeed a factor $\sim 3$ higher than the limiting flux of
the BGS (5.24 Jy). The sources are listed in Table 2.

\begin{table}
\centerline{\begin{tabular}{cccc}
\hline
NAME & F$_{IR}^1$& X-ray$^2$ & Class.$^3$ \\
\hline
NGC 1068        & 29  & Y & AGN \\
NGC 3690        & 11  & Y & ST. \\
NGC 1365        & 9.9 & Y & AGN \\
NGC 3256        & 9.1 & Y & ST. \\
ARP 220         & 7.7 & Y & ST. \\
NGC 1614        & 3.3 & Y & AGN \\
IRAS 18293-3413 & 3.2 & Y & AGN \\
NGC 7469        & 2.7 & Y & AGN \\
ESO 320-G030    & 2.7 & N & AGN \\
MKN 231         & 3.1 & Y & AGN \\
IRAS 17208-0014 & 2.5 & Y & ST. \\
IC 1623         & 2.1 & N  & ST. \\
MCG+12-02-001   & 2.1 & N & ST. \\
IC 5179         & 2.1 & N & ST. \\
IC 4687         & 2.1 & N & ST. \\
NGC 2369        & 2.1 & N & ST. \\
NGC 6240        & 2.0 & Y & AGN \\
NGC 7771        & 2.0 & N & ST. \\
NGC 2623        & 1.8 & Y & AGN \\
NGC 5135        & 1.7 & Y & AGN \\
MKN 273         & 1.7 & Y & AGN \\
ZW 049.057      & 1.6 & N & AGN \\
NGC 7130        & 1.6 & Y & AGN \\
\hline
\end{tabular}}
\caption{\footnotesize{An unbiased sample of LIGs extracted from the BGS
catalogue. Notes: $^1$ Total IR flux in units of 10$^{-9}$ erg cm$^{-2}$
s$^{-1}$; $^2$ Observed at 2-10 keV; $^3$ Optical classification.}}
\end{table}

Although a complete statistical study of their properties
is at present impossible because of the lack of X-ray data for 8
sources, we can nonetheless draw some preliminary conclusions. According to 
the optical classification, 5 of the 15 X-ray observed sources are starbursts,
while the remaining 10 are AGNs. Among the latter, 6 are Compton--thick and
only one (NGC 1365) is Compton--thin with a column density $\rm N_H \sim
10^{23}$ cm$^{-2}$. In three sources, the AGN seen in the optical is
completely invisible in the X rays.
Finally, the 25/60 $\mu$m flux 
ratio is generally low, and indicates that the emission of 
the warm dust associated with the AGN is either low compared with the 
starburst 60$\mu$m emission, or absorbed.

The statement we can make is that the AGNs are
rather common among the luminous infrared galaxies 
(13 out of 23 in the whole sample), but their contribution
to the bolometric luminosity is not easy to assess. The AGN could
provide the higher fraction of the energy only if its direct emission were
entirely reprocessed; alternatively, the contribution of the starburst
could be important or even dominant.

\section{Conclusions}

This paper deals with the sample of luminous infrared galaxies (LIGs) 
observed to date in hard X~rays (2-10 keV).
The sample is affected by a selection bias in favor of AGN-dominated
sources, nevertheless it covers a region of the parameter space large
enough to allow some general conclusions.
For a significant fraction of the sample (15/78) hard X-ray data
are analyzed and published for the first time.

The main results of our work can be summarized as follows:
\medskip

\noindent
1) The X-ray properties of the LIGs are very different from source to
source: the X-ray brightest objects are Seyfert~1 and low--absorption
Seyfert~2 galaxies, with a X/IR flux ratio as high as 0.1-0.2.
At the opposite extreme are the starburst dominated objects, for
which the X/IR flux ratio is $\sim 10^{-4}$ or lower. Interestingly, we
note that a significant fraction of the sources optically classified as
AGNs do not show any indication of nuclear activity in X~rays.
Thus, either the AGN contribution is negligible, or the direct emission 
must be absorbed by a column density
$\rm N_H > 10^{24}$ cm$^{-2}$. 
\medskip

\noindent
2) A subsample of 13 LIGs was
observed up to 200 keV. Six out of 11 sources that are
Compton--thick in the 2-10 keV range are reflection dominated also from
10 to 200 keV, thus implying a column density $\rm N_H>10^{25}$ cm$^{-2}$
(or, alternatively, a fading nucleus).
Only two have an excess in the 15-100
keV range, while for the last three the hard 15-100 keV X-ray emission remains
unconstrained.
This result is somewhat puzzling, for the 10$^{24} - 10^{25}$ cm$^{-2}$
interval of column density appears to be underpopulated both for the LIGs
and the non-LIG local Seyfert~2 galaxies.
\medskip

\noindent
3) The shape of the correlation between the X/IR flux ratio and the
25/60 $\mu$m IR colour suggests that the 25$\mu$m emission is absorbed
by a lower column density than the X-ray emission or, alternatively, 
that the absorbing material has a dust--to--gas ratio lower than Galactic.

The X-IR correlation is well reproduced by a model in which both AGNs
and starbursts contribute to the total emission. The differences in the
IR colours are mainly due to the different contribution of the
starburst component, while the X/IR flux ratio is mainly determined 
by the amount of absorption affecting the AGN.

Additional evidence in support of the model is the detection of broad
optical polarized or infrared lines only in the IR-warmest sources 
(where we see directly the inner region of the torus), and the steepness 
of the X-ray spectra of the IR-colder objects (where the starburst 
contributes strongly).
\medskip

\noindent
4) Finally, we assembled an unbiased sample of LIGs from the BGS catalog,
in a way which maximizes the fraction of sources with hard X-ray data
available (15 out of 23, i.e. 65\%). The sample is limited in {\it total}
IRAS flux (i.e. from 12$\mu$m to 100$\mu$m). This selection criterion
removes a bias against AGNs present in most other LIG samples
that, instead, are limited in the 60$\mu$m flux.
 From the optical classification
and from the IR and X-ray data we find that AGNs are common among the
LIGs (13 out of 23 host an active nucleus) but they are
either weak or heavily obscured.

\begin{acknowledgements}
Three of us (RG, RM and MS) were supported in part by the Italian Ministry for
University and Research (MURST) through grant Cofin 98-02-32. The detailed
comments of the referee, K. Iwasawa, have greatly improved the presentation.
\end{acknowledgements}

\appendix
\section{Unpublished sources}

We present here a brief discussion of the X-ray data of 15 sources of our
sample for which no information is available in the literature.
Among the 10 sources observed by ASCA, only 3 (TOL 1351-375, IRAS
05189-2524 and MKN 1048) have a good signal-to-noise. For them a detailed study 
is possible, and the main spectral parameters can be determined. 
The remaining 7 sources (3 classified as starbursts and 4 as AGNs) have
detections at a few sigmas and therefore we can only estimate a flux and
sometimes a photon index.
The 5 sources observed by BeppoSAX are all very faint in the
X-rays, and for 4 of them the X-ray observation does not provide any firm
indication of the presence of an AGN. For the fifth object, MKN 266,
the detection of a cold iron line and the flatness of the spectrum suggest
the presence of a heavily obscured AGN.
All the errors quoted are at the 90\% level of confidence.
\medskip

\noindent
{\it a) Sources from the ASCA public archive:}\hfill\break
We have reduced and analyzed the ASCA GIS observations of 11 sources. All 
the data have been retrieved from the ASCA public archive.\hfill\break
{\bf IRAS 03158+4227:}
The GIS data of IRAS 03158+4227 give a detection at a low
signal--to--noise level ($\sim 6 \sigma$). The spectrum is well fitted
by a powerlaw with $\Gamma \simeq 2.8$. The 2-10 keV flux is F=5.4$\pm1
\times 10^{-13}$ erg cm$^{-2}$ s$^{-1}$. There is no evidence of the
iron line, and the steepness of the spectrum suggests a thermal origin
of the observed emission.
IRAS 03158+4227 is classified optically as Seyfert 2, therefore it should
host an AGN. From the low X-ray luminosity, relative to the
infrared emission, and from the absence of AGN indicators in the X~rays
we deduce that the AGN is obscured by a column density $\rm N_H > 10^{24}$
cm$^{-2}$.\hfill\break
{\bf IRAS 07598+6508:}
The source IRAS 07598+6508 has not been detected in the GIS observation
with an exposure time of 83500 sec. We can only derive an upper limit 
to the 2-10 keV flux, F$^{MAX}$=8 10$^{-14}$ erg cm$^{-2}$ s$^{-1}$
(at a 90\% confidence level).\hfill\break
{\bf IRAS 17208-0014:} 
The source IRAS 17208-0014 is marginally detected in the GIS.
The flux we derive is F=3$\pm1 \times 10^{-13}$ erg cm$^{-2}$ 
s$^{-1}$.\hfill\break
{\bf TOL 1351-375:}
TOL 1351-375 is a Seyfert 1.9 galaxy observed by ASCA in 1997. The
spectrum obtained after the standard data reduction has a good 
signal-to-noise and has been fitted with a multi-component model consisting 
of an absorbed powerlaw, an iron K$_\alpha$ line and a thermal component
with kT$\simeq$1.2 keV. The best fit gives the following values: photon
index $\Gamma=1.94\pm0.13$; absorbing column density $\rm N_H=1.6\pm 3
\times 10^{22}$ cm$^{-2}$;
rest frame line energy E=6.38$^{+0.2}_{-0.3}$ keV; line equivalent width
EW=175$\pm$120 eV. The measured 2-10 keV flux is 3.8 $10^{-12}$ erg
cm$^{-2}$ s$^{-1}$.\hfill\break
{\bf IRAS 18293-3413:}
This source has been detected at a 5$\sigma$ level. There is no evidence
of an AGN contribution. Fitting the data with a blackbody we obtain
kT=0.85 keV. The 2-10 keV flux is 9 $10^{-14}$ erg cm$^{-2}$ 
s$^{-1}$.\hfill\break
{\bf NGC 1614:} 
The spectrum of NGC 1614 is well fitted by a powerlaw with $\Gamma=
1.55\pm0.4$, while a termal component is rejected. There is no evidence
of an iron line. The measured 2-10 keV flux is 5.6 $10^{-13}$ erg
cm$^{-2}$ s$^{-1}$.\hfill\break
{\bf IRAS 05189-2524:} 
This source was observed by ASCA for $\sim$ 40000 seconds.
The spectrum is well fitted by a two-component model, consisting of a
powerlaw with photon index $\Gamma = 1.89^{+0.35}_{-0.34}$ absorbed by a
column density of $\rm N_H = 4.7^{+1.4}_{-1.1} \times 10^{22}$ cm$^{-2}$,
and a thermal component with kT=0.88$^{+0.89}_{-0.35}$. The iron line is
not detected. From the non-detection we estimate an upper limit to the
equivalent width of 235 eV. The measured 2-10 keV flux is 5.3$\times
10^{-12}$ erg cm$^{-2}$ s$^{-1}$.
In summary, the E$>2$ keV spectrum is typical of a Compton--thin type 2
AGN, while at lower energies an extra thermal component is present, that
can be either associated to the AGN itself or due to starburst
activity.\hfill\break
{\bf NGC 7212:}
The Seyfert 2 galaxy NGC 7212 has a flat 0.5-10 keV spectrum, well
fitted by a powerlaw with photon index $\Gamma = 0.75^{+0.50}_{-0.55}$.
An emission line with peak energy E=6.06$^{+0.35}_{-0.40}$ (compatible
with a K$_\alpha$ iron line at a confidence level of 90\%) is detected
at a 2$\sigma$ level, with a best fit equivalent width
EW=1.0$^{+1.3}_{-0.9}$ keV. From these spectral features we deduce that
the source is Compton-thick. The measured 2-10 keV flux is F=9.8$\times
10^{-13}$ erg cm$^{-2}$ s$^{-1}$.\hfill\break
{\bf MKN 1048:}
The GIS spectrum of MKN 1048 is well fitted by a simple powerlaw with
photon index $\Gamma=1.60^{+0.02}_{-0.01}$, a value significantly lower
than the average photon index found in the X-ray spectra of quasars.
The K$_\alpha$ iron line is marginally detected, at a 2$\sigma$ level
(best fit equivalent width EW=96$_{-83}^{+86}$ eV.
The measured 2-10 keV flux is F=9.9$\times 
10^{-12}$ erg cm$^{-2}$ s$^{-1}$.\hfill\break
{\bf WAS 49b}
The Seyfert 2 galaxy WAS 49b presents a very unusual X-ray spectrum.
A  good analytical fit is provided by a single powerlaw with photon index
 $\Gamma =0.5$, a very low value typical of reflection-dominated spectra.
In contrast with this interpretation, the iron line is only marginally
detected with an equivalent width EW $\simeq 250\pm 200$ eV and the 2-10 keV
flux is relatively high with respect to the infrared emission 
(F$_{2-10}= 1.2 \times 10^{-12}$ erg cm$^{-2}$
s$^{-1}$), thus indicating that the source is Compton - thin. 
A similar good fit is obtained by a model composed by a powerlaw with
$\Gamma =1.7$ absorbed by a column density $\rm N_H \sim 10^{23}$ cm$^{-2}$ and
a flat ($\Gamma =0.6$) powerlaw with a 1 keV normalization equal to 1/6 that
of the first component. Since WAS 49b is part of a triple sistem, we suggest 
that this flat contribution is due to the
diffuse emission of the hot intra-cluster gas.   

\medskip

\noindent
{\it b) Sources observed with BeppoSAX:}\hfill\break
{\bf MKN 273:} 
MKN 273 was observed with BeppoSAX in 1998 and was detected by the MECS
(1.65-10 keV) at a 2-10 keV flux of 3.5$\times 10^{-13}$ erg
cm$^{-2}$ s$^{-1}$, significantly lower than the flux measured by ASCA 2
years before (F$_{2-10}=7\times 10^{-13}$ erg
cm$^{-2}$ s$^{-1}$, Iwasawa 1998). Fitting the data with a powerlaw and
an iron line we obtain a spectral index ($\Gamma=1.1_{-0.6}^{+0.9}$,
poorly constrained due to the weakness of the continuum, and a
marginal detection of the iron line, at a 3$\sigma$ level (best fit
equivalent width EW=1.2$_{-1}^{+2}$ keV). The detection in the PDS is
also weak. We conclude that either 
the AGN at the centre of this source is absorbed 
by a column density $\rm N_H > 10^{25}$ cm$^{-2}$ or, alternatively, that the
central source has faded and only the reflected component is now
visible (this would also explain the flux decrease between the ASCA
and SAX observations).\hfill\break
{\bf MKN 266:} 
MKN 266 was detected by the MECS and
marginally (at a 2$\sigma$ level) by the PDS.
The 2-10 keV spectrum is remarkably flat (photon index $\Gamma =
0.7^{+0.4}_{-0.3}$) suggesting the presence of a heavily obscured AGN.
The iron line is detected at a 90\% confidence level, with a best fit
equivalent width of 575 eV. The absence of any excess in the PDS is an
indication of a column density higher than 10$^{25}$ cm$^{-2}$.
The measured 2-10 keV flux is 5.6 $10^{-13}$ erg
cm$^{-2}$ s$^{-1}$.\hfill\break
{\bf NGC 2623:} 
NGC 2623 was detected by the MECS at 3.5$\sigma$, thus we can only
estimate the 2-10 keV flux, F$_{2-10}\simeq 8\times 10^{-14}$ erg
cm$^{-2}$ s$^{-1}$. The detection in the PDS is weak and cannot be
associated with certainty to NGC 2623, because of the presence in the PDS
field ($\sim 100$) of another source that could emit in the hard
X~rays.\hfill\break
{\bf MKN 1073:}
MKN 1073 was not detected by the MECS and therefore we can only estimate
an upper limit to the 2-10 keV flux, F$_{2-10} > 7 10^{-14}$ erg
cm$^{-2}$ s$^{-1}$. The signal in the PDS is very high, with an
exceptionally steep spectrum ($\Gamma \sim 3$), but it is
probably emitted by an unidentified source in a nearby cluster, that has 
a strong emission in the 2-10 keV band as well.\hfill\break
{\bf IRAS 00198-7926}
The source IRAS 00198-7926 was observed by the BeppoSAX instruments for
$\sim$ 20000 seconds and was not detected. From this non-detection we
estimate an upper limit to the 2-10 keV flux, F$_{2-10} > 1\times 10^{-13}$ erg
cm$^{-2}$ s$^{-1}$.

\end{document}